# Vanishing skyrmion Hall effect at the angular momentum compensation temperature of a ferrimagnet

Yuushou Hirata[1], Duck-Ho Kim[1]★, Se Kwon Kim[2], Dong-Kyu Lee[3], Se-Hyeok Oh[4], Dae-Yun Kim[5], Tomoe Nishimura[1], Takaya Okuno[1], Yasuhiro Futakawa[6], Hiroki Yoshikawa[6], Arata Tsukamoto[6], Yaroslav Tserkovnyak[2], Yoichi Shiota[1], Takahiro Moriyama[1], Sug-Bong Choe[5], Kyung-Jin Lee[3,4,7]★, and Teruo Ono[1,8]★

[1]Institute for Chemical Research, Kyoto University, Uji, Kyoto 611-0011, Japan

[2]Department of Physics and Astronomy, University of California Los Angeles, California 90095, USA

[3]Department of Materials Science & Engineering, Korea University, Seoul 02841, Republic of Korea

[4]Department of Nano-Semiconductor and Engineering, Korea University, Seoul 02841, Republic of Korea

[5]Department of Physics and Institute of Applied Physics, Seoul National University, Seoul 08826, Republic of Korea

[6]College of Science and Technology, Nihon University, Funabashi, Chiba 274-8501, Japan

[7]KU-KIST Graduate School of Converging Science and Technology, Korea University, Seoul 02841, Republic of Korea

[8]Center for Spintronics Research Network (CSRN), Graduate School of Engineering Science, Osaka University, Osaka 560-8531, Japan

★Corresponding authors. E-mails: uzes13@gmail.com (D.-H. Kim), kj_lee@korea.ac.kr (K.-J. Lee), ono@scl.kyoto-u.ac.jp (T. Ono)

**Charged particles exhibit the Hall effect in the presence of magnetic fields. Analogously, ferromagnetic skyrmions with non-zero topological charges and finite fictitious magnetic fields exhibit the skyrmion Hall effect, which is detrimental for applications. The skyrmion Hall effect has been theoretically predicted to vanish for antiferromagnetic skyrmions because the fictitious magnetic field, proportional to net spin density, is zero. We experimentally confirm this prediction by observing current-driven transverse elongation of pinned ferrimagnetic bubbles. Remarkably, the skyrmion Hall effect, estimated with the angle between the current and bubble elongation directions, vanishes at the angular momentum compensation temperature where the net spin density vanishes. This study establishes a direct connection between the fictitious magnetic field and spin density, offering a pathway towards the realization of skyrmionic devices.**

The skyrmion[1] is a topological field configuration with particle-like properties that appears in wide ranges from microscopic to cosmological scales[2,3]. One particular form of skyrmions is the magnetic skyrmion, found in chiral magnets[4,5]. The magnetic skyrmion offers an experimental framework for investigating topological effects on soliton dynamics. Its representative topological effect is the skyrmion Hall effect[6–8], analogous to the Hall effect for which the Lorentz force deflects charged particles perpendicular to their velocities.

Magnetic skyrmion dynamics is described by the equation of motion for position $\bm{R}$ [9–11], given as:

$$M\ddot{\bm{R}} = Q\dot{\bm{R}} \times \bm{B} - D\dot{\bm{R}} + \bm{F}, \qquad (1)$$

where $M$ is the soliton mass, $D$ is the viscous coefficient, and $\bm{F}$ is the sum of the internal force originating from the potential energy and external driving force. The skyrmion Hall effect results from the first term on the right-hand side, an effective Lorentz or Magnus force, defined with topological charge $Q \equiv \int dxdy\, \boldsymbol{n} \cdot (\partial_x \boldsymbol{n} \times \partial_y \boldsymbol{n})/4\pi$ and fictitious magnetic field $\boldsymbol{B} = -4\pi s_{net} \hat{\boldsymbol{z}}$, where $\boldsymbol{n}$ is the spin order parameter and $s_{net}$ is the net spin density. The skyrmion Hall effect is thus determined by the topological charge $Q$ and fictitious magnetic field $\boldsymbol{B}$, similar to the Hall effect determined by the electric charge and external field.

For ferromagnetic skyrmions, both topological charge and spin density are finite so that the skyrmion Hall effect emerges[6,7]. The topological charge remains finite even for antiferromagnetic skyrmions because $\boldsymbol{n}$ is defined by the Néel vector (not by alternating atomic spin), which is continuous in space. Thus, theoretically predicted vanishing skyrmion Hall effect for antiferromagnetic skyrmions[12–14] is entirely associated with zero net spin density or, equivalently, zero fictitious magnetic field. This theoretical prediction has not yet been experimentally verified.

The significance of experimentally demonstrating the vanishing skyrmion Hall effect is twofold. First, it is crucial for realizing spintronic devices that utilize magnetic skyrmions as information carriers[15–23], because the skyrmion Hall effect pushes skyrmions towards edges of patterned devices, potentially resulting in skyrmion annihilation and loss of information. Furthermore, antiferromagnetic spin textures are expected to move faster than ferromagnetic ones[12–14,24–26], which is beneficial for energy-efficient devices. Second, although the fictitious magnetic field of topological spin textures – critical for understanding their dynamics – has been predicted to be proportional to the spin density[9–11,14], a direct connection between the fictitious magnetic field and spin density has remained elusive. This connection can be established by independent measurements of the vanishing skyrmion Hall effect and zero spin density, as shown below.

In this work, we experimentally demonstrate the vanishing skyrmion Hall effect using a rare-earth (RE=Gd) and transition-metal (TM=FeCo) ferrimagnetic compound where RE and TM moments are antiferromagnetically coupled. RE and TM elements have different intra-atomic exchange and thus exhibit different temperature-dependent spin density changes. As a result, the net spin density of GdFeCo ferrimagnets varies gradually with temperature and vanishes at a specific temperature below the Curie temperature. The nature of magnetic dynamics, which is governed by the angular momentum and their commutation relations, becomes antiferromagnetic at this temperature, called the angular momentum compensation temperature $T_A$[27]. This feature of GdFeCo ferrimagnets allows us to experimentally test the relation of the fictitious magnetic field to the spin density, and in particular, the vanishing skyrmion Hall effect at $T_A$.

We fabricate perpendicularly magnetized ferrimagnetic GdFeCo/Pt films[28] via sputtering. The Pt layer serves as a spin-current source, and an in-plane charge current generates a spin-orbit torque (SOT)[29,30]. This layered structure has a sizable Dzyaloshinskii-Moriya interaction (DMI), larger than domain-wall (DW) hard-axis anisotropy[28], and thus homochiral magnetic bubbles with well-defined topological charges can be stabilized in the structure. We first determine $T_A$ of the GdFeCo/Pt film by measuring the field-driven DW velocity[27]. Figure 1**a** shows the DW velocity $v$ as a function of the temperature $T$ under a perpendicular magnetic field, $\mu_0 H_z = 50$ mT. It clearly demonstrates a sharp peak in $v$ at $T =287\pm5$ K, corresponding to $T_A$[27]. This drastic increase in $v$ evidences antiferromagnetic spin dynamics, corresponding to zero net spin density at $T_A$.

Figure 1**b** shows an optical microscope image of the device used in the SOT experiment. To create a bubble domain, we apply a current pulse along the y-axis through a half-ring shaped writing line, prepared via electron beam lithography (inset of Fig. 1**b**). The current-induced Oersted field creates a reversed bubble domain predominantly at the location

of the half ring and occasionally bubble domains near the edge of the writing line, possibly due to lithography-induced damage. Figure 2**a** shows a magneto-optical Kerr effect (MOKE) image of a created bubble domain (magnetization-UP state) at the half ring. Lighter and darker areas correspond to regions of magnetization-UP and magnetization-DOWN states, respectively. We then inject a current pulse (amplitude $I = 0.14$ A and duration $\Delta t = 100$ ms) through the GdFeCo/Pt film along the x-axis at $T = 343$ K ($T > T_A$) for which the current-induced Joule heating effect is corrected[28]. This in-plane current generates a SOT that could move the bubble domain. As shown in Fig. 2**b**, however, we observe that the current elongates the bubble domain, instead of moving it, possibly due to lithography-induced damage that creates a strong pinning potential.

This current-driven bubble domain elongation however does not alter the main conclusion of this research, as explained below. In Fig. 2**b**, the bubble elongation direction (indicated by a red dotted line) has an angle $\theta \approx -35°$ with respect to the current direction. To investigate the physical meaning of this angle, we perform the same experiment for a bubble domain with the opposite magnetization (magnetization-DOWN; Fig. 2**c**). We find that $\theta$ in this case is about $+31°$ (Fig. 2**d**). Thus, changing the bubble domain magnetization state from UP to DOWN changes the sign of $\theta$ while approximately maintaining its absolute value. This sign change, due to the overall inversion from the spin order parameter $\boldsymbol{n}$ to $-\boldsymbol{n}$, suggests that the elongation angle $\theta$ reflects the topological charge $Q$ because $Q$ is the odd function of $\boldsymbol{n}$.

The correlation between elongation angle and topological charge in this experiment indicates that the current-driven elongation of bubble domains is a consequence of half-skyrmion motion (Fig. 2**e**). As one bubble edge is pinned, the motion of the other edge, indicated by a dotted box in Fig. 2**e**, is responsible for elongation. For a bubble domain having a well-defined topology due to the DMI, a topological charge for the other edge in the boxed area is a half a full skyrmion and, as a result, the elongation angle could be interpreted as the skyrmion Hall effect of a half-skyrmion.

To validate the above interpretation, we investigate current-driven elongation of bubble domains at various temperatures. If the above interpretation is valid, $\theta$ must vary gradually with $T$ and change its sign at $T_A$ because the skyrmion Hall effect for a bubble domain with a well-defined $Q$ is determined by the net spin density. Figures 3**a-c** show MOKE images of the magnetization-DOWN state, obtained at three representative temperatures. The sign of $\theta$ is positive above $T_A$ ($T = 343$ K; Fig. 3**a**) whereas it is negative below $T_A$ ($T = 253$ K; Fig. 3**c**). Remarkably, $\theta$ is close to zero at the temperature near $T_A$ ($T = 283$ K; Fig. 3**b**). Figure 3**d** summarizes the measured $\theta$ as a function of $T$, revealing that bubble domains with magnetization-UP and magnetization-DOWN states exhibit $\theta$ of opposite signs and comparable absolute values throughout the tested temperature range. More importantly, the $\theta$ for both types of bubble domain approaches zero at temperature close to $T_A$. This observation is consistent with the above interpretation based on half skyrmions, demonstrating that the skyrmion Hall effect of RE-TM ferrimagnets vanishes at $T_A$.

To further support the proposed interpretation with experimental observations, we develop a simple theory for the elongation of a pinned magnetic bubble driven by SOT. Specifically, we model an elongated bubble with one pinned end at the origin (Fig. 4**a**) as a composite object consisting of a half skyrmion at the free end and a straight rod connecting the two ends, which resembles the model used in Ref. 19 to explain skyrmion generation from the tips of elongated bubbles. The state of the bubble can be described by two variables, the rod length $l(t)$ and the elongation angle from the current direction $\theta(t)$. Their equations of motion can be obtained using the collective coordinate approach[11,14,27], the details of which can be found in Ref. 28. Several forces act on the half skyrmion: the SOT-induced force $\boldsymbol{F}_j$; the Magnus force $\boldsymbol{F}_g$, proportional both to the spin density and the topological charge $Q = \pm 1/2$; the viscous force $\boldsymbol{F}_d$, rooted in the Gilbert damping, and the tension $\boldsymbol{F}_T$, associated with the stretching of the rod (see Fig. 4a for schematic illustrations of these forces). By balancing these forces and the associated torques on the bubble, we obtain the steady-state solution with uniform growth and a constant angle for sufficiently large currents: $\dot{l}(t) = v$ and $\theta(t) = \theta_{\text{SkH}}$ with

$$\theta_{\text{SkH}} \approx \tan^{-1}\left(\frac{2s_{\text{net}} Q \lambda}{\alpha s_{\text{total}} r}\right), \qquad (2)$$

where $s_{\text{total}}$ is the sum of the sublattice spin density magnitudes, $\alpha$ is the Gilbert damping constant, $\lambda$ is the DW width that forms the bubble boundary, and $r$ is the radius of the half skyrmion. The elongation angle obtained for the bubble is identical to the expression for the skyrmion Hall angle for an isolated skyrmion[14]. As the temperature approaches $T_A$, the net spin density $s_{\text{net}}$ goes to zero and, consequently, the elongation angle $\theta_{\text{SkH}}$ goes to zero. This implies the skyrmion Hall effect vanishes at $T_A$, as was experimentally observed.

We also perform numerical simulations based on the atomistic Landau-Lifshitz-Gilbert (LLG) equation[27,28]. We compute SOT-driven elongation of a ferrimagnetic bubble by varying the spin densities of two sub-lattices around $s_{\text{net}} = 0$ [28]. To mimic the experiment, we intentionally pin one bubble edge by applying a staggered field locally and assume a random distribution of the perpendicular magnetic anisotropy[28]. Figure 4b shows snapshot images of the elongated magnetic bubble for nine $s_{\text{net}}$ cases. Figure 4c summarizes the numerically obtained elongation angles as a function of $s_{\text{net}}$. At $s_{\text{net}} = 0$, the elongation angle is nearly zero, consistent with both experiment and theory.

We experimentally observe that the skyrmion Hall effect in an antiferromagnetically coupled ferrimagnet varies with temperature and vanishes at $T_\text{A}$. Combined with theoretical and numerical support, this observation demonstrates that the fictitious magnetic field for magnetic skyrmions is proportional to the net spin density, revealing the importance of the net spin density in topological soliton dynamics where it has been overlooked compared to the topological charge. Furthermore, this demonstration of the vanishing skyrmion Hall effect is of crucial importance for realizing an efficient skyrmion racetrack memory without information loss. The high tunability of the net spin density of ferrimagnets, which cannot be found in more conventional spintronic materials, will open up new possibilities for manipulating the dynamics of topological solitons and thereby facilitate the realization of soliton-based spintronic devices.

## Acknowledgements

This work was supported by the JSPS KAKENHI (Grant Numbers 15H05702, 26103002, and 26103004), Collaborative Research Program of the Institute for Chemical Research, Kyoto University, and R & D project for ICT Key Technology of MEXT from the Japan Society for the Promotion of Science (JSPS). This work was partly supported by the Cooperative Research Project Program of the Research Institute of Electrical Communication, Tohoku University. D.H.K. was supported as an Overseas Researcher under the Postdoctoral Fellowship of JSPS (Grant Number P16314). D.K.L., S.H.O., and K.J.L. were supported by the National Research Foundation of Korea (NRF-2015M3D1A1070465, 2017R1A2B2006119) and the KIST Institutional Program (Project No. 2V05750). S.K.K. and Y.T. were supported by the Army Research Office under Contract No. W911NF-14-1-0016. D.Y.K. and S.B.C. were supported by a National Research Foundations of Korea (NRF) grant funded by the Ministry of Science, ICT and Future Planning of Korea (MSIP) (2015R1A2A1A05001698 and 2015M3D1A1070465).

## Author contributions

K.-J.L., D.-H.K., and T.O. planned and designed the experiment. Y.F., H.Y., and A.T. prepared GdFeCo ferrimagnetic films and Y.H. prepared the devices. Y.H., D.-H.K., D.-Y.K., and T.N. carried out the measurement. S.K.K., K.-J.L. and Y.T. provided theory. D.-H.K. and Y.H. performed the analysis of experimental results. D.-K.L., S.-H.O., and K.-J.L. performed the numerical simulation. D.-H.K., K.-J.L., S.K.K., Y.H., S.-B.C. and T.O. wrote the manuscript. All authors discussed the results and commented on the manuscript.

## Competing interests

Authors declare no competing interests.

## Data and materials availability

All data are available in the main text or the supplementary information.

**Figure captions**

**Figure. 1. Determination of $T_A$ and device structure. a**, $v$ as a function of $T$ under $\mu_0 H_z$ =50 mT, with arrow indicating $T_A$. **b**, Optical image of device structure and definition of the coordinate system.

**Figure. 2 Determination of current-driven elongation of pinned magnetic bubbles at $T$ =343 K ($T > T_A$). a**, MOKE image of magnetization-UP bubble domain. **b**, Current-driven elongation of magnetic bubble from **a**. **c**, MOKE image of a magnetization-DOWN bubble domain. **d**, Current-driven elongation of magnetic bubble from **c**. **e**, Illustration of current-driven elongation of magnetic bubble. Yellow arrows indicate magnetization directions at the DW section. An edge surrounded by a dotted box is responsible for elongation while the other edge is pinned. The angle $\theta$ in (**b**, **d**, and **e**) is the elongation angle.

**Figure. 3. Elongation angle $\theta$ as a function of $T$.** MOKE images at **a**, $T$ =343 K ($T > T_A$), **b**, $T$ =283 K ($T \approx T_A$), and **c**, $T$ =253 K ($T < T_A$). **d**, $\theta$ as a function of $T$ for each magnetization state.

**Figure. 4. Theoretical and numerical results for current-driven elongation of pinned ferrimagnet bubble. a**, Forces acting on a half skyrmion represented by the grey solid circle (see the main text for each force type). **b**, Numerical results for the SOT-driven elongation of a pinned magnetic bubble for $s_{net}$[28]. $\mathbf{n}_z$ is $(\mathbf{m}_{i+1} - \mathbf{m}_i)/2$, where $\mathbf{m}_i$ is the normalized magnetic moment at atomic site $i$[28]. **c**, Numerically obtained $\theta$ as a function of $s_{net}$ for $Q = \pm 1/2$.

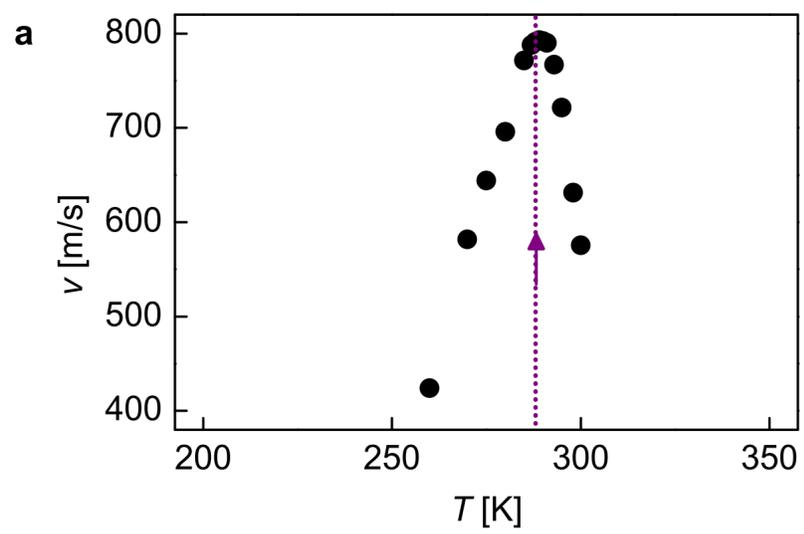

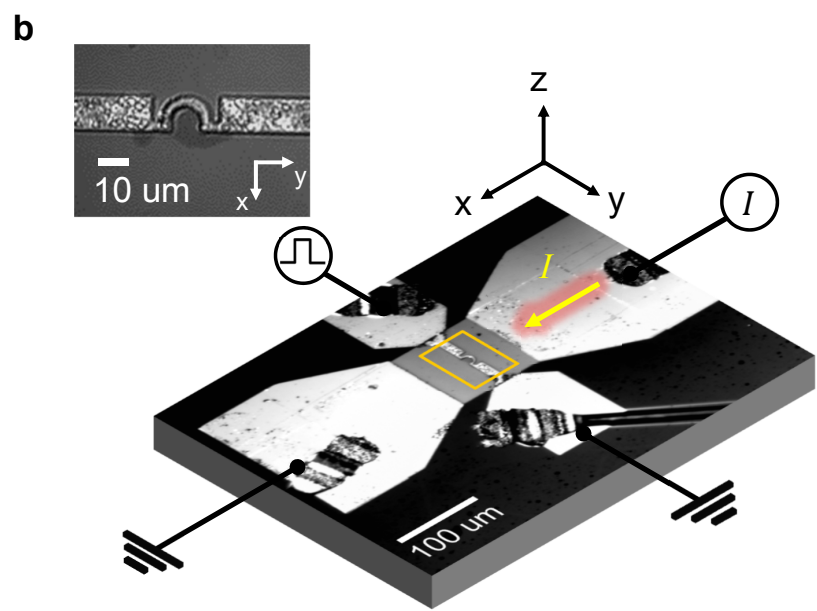

Figure 1

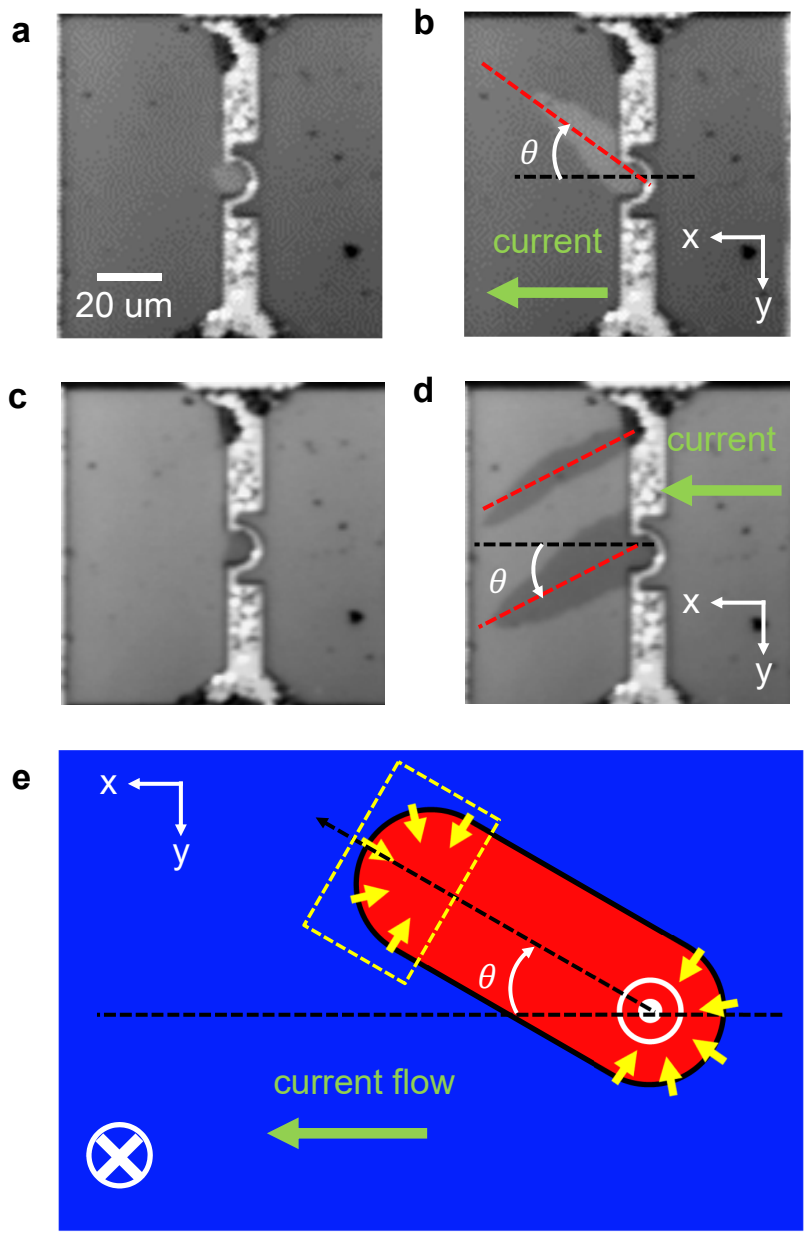

Figure 2

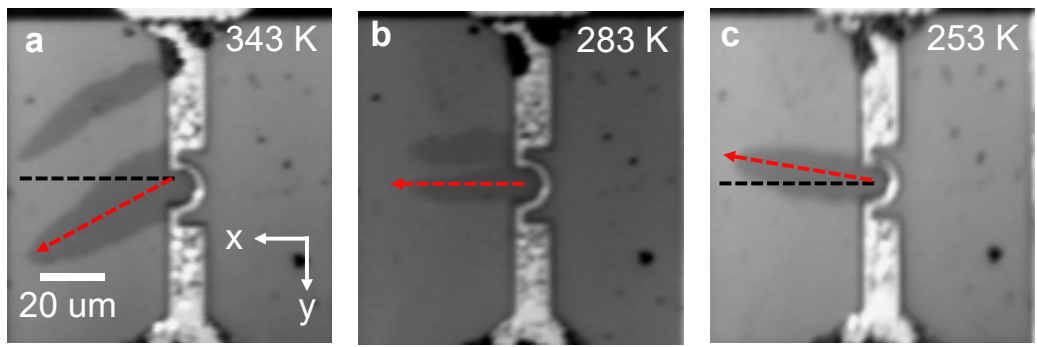
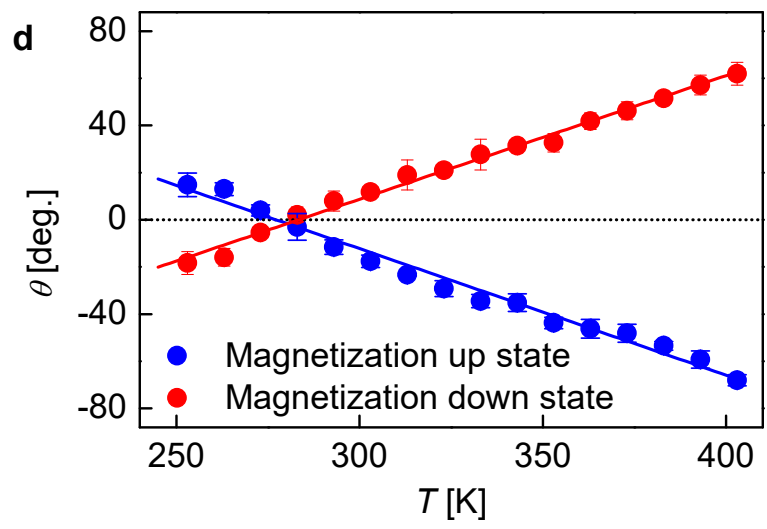

Figure 3

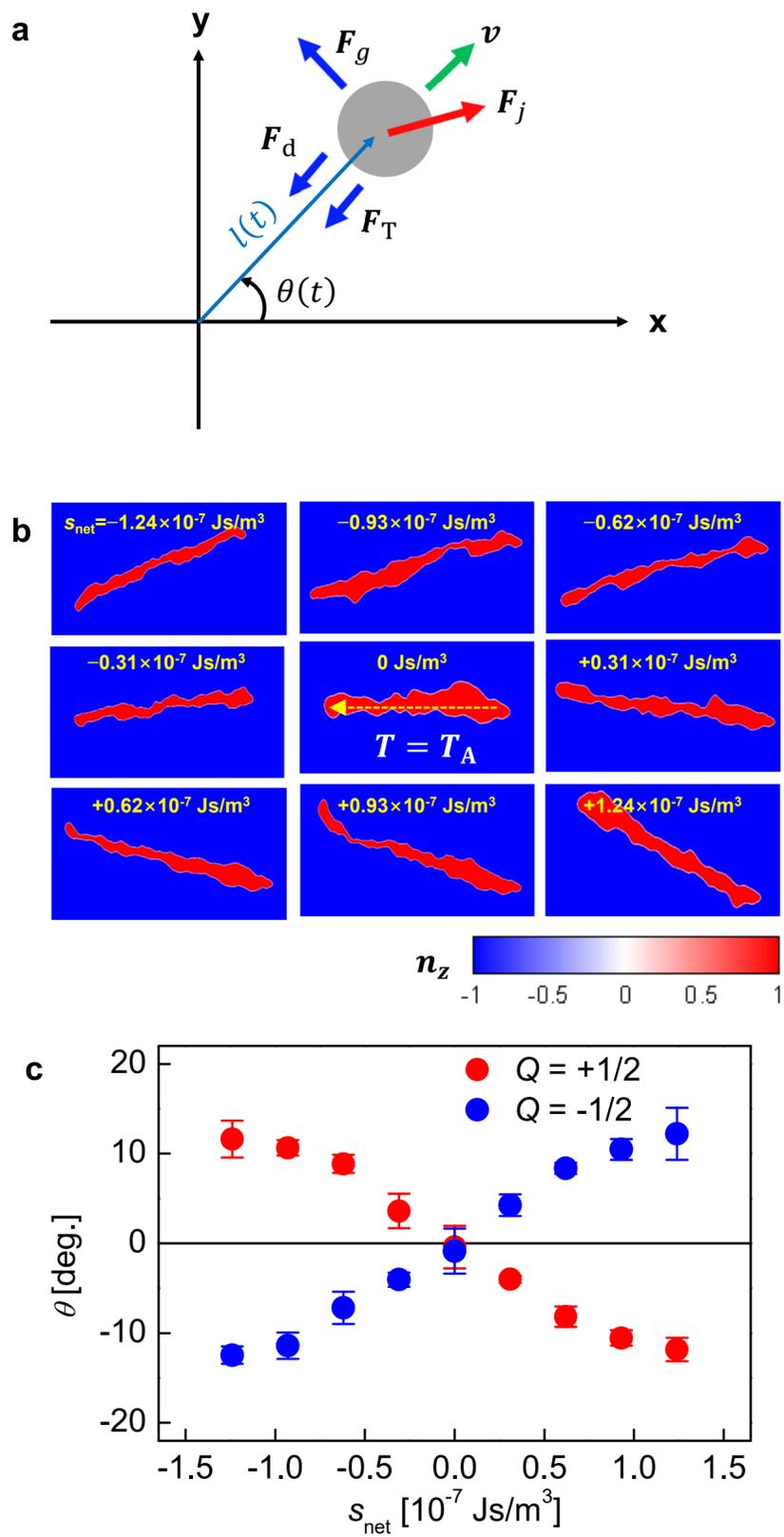

Figure 4